\documentclass{article}

\usepackage[dvips]{graphics,graphicx,color}
\usepackage{a4,indentfirst,latexsym,amsfonts,amsmath,amssymb,array}
\usepackage{wrapfig}

\begin{document}

\title{The influence of initial impurities and irradiation conditions on defect production 
and annealing in silicon for particle detectors}

\author {Ionel Lazanu\footnote{University of Bucharest, POBox MG-11, Bucharest-Magurele, 
Romania} \hspace{2 pt}
and Sorina Lazanu\footnote{National Institute for Materials Physics, POBox MG-7, 
Bucharest-Magurele, Romania}}

\date{\today}
\maketitle
\begin{abstract}
Silicon detectors in particle physics experiments at the new accelerators or in space 
missions for physics goals will be exposed to extreme radiation conditions. The principal 
obstacles to long-term operation in these environments are the changes in detector 
parameters, consequence of the modifications in material properties after irradiation.

The phenomenological model developed in the present paper is able to explain 
quantitatively, without free parameters, the production of primary defects in silicon 
after particle irradiation and their evolution toward equilibrium, for a large range of 
generation rates of primary defects. Vacancy-interstitial annihilation, interstitial 
migration to sinks, divacancy and vacancy-impurity complex ($VP$, $VO$, $V_2O$, $C_iO_i$ 
and $C_iC_s$) formation are taken into account. The effects of different initial impurity 
concentrations of phosphorus, oxygen and carbon, as well as of irradiation conditions are 
systematically studied. The correlation between the rate of defect production, the 
temperature and the time evolution of defect concentrations is also investigated.

61.80 Az: Theory and models of radiation defects\\
61.70 At: Defect formation and annealing\\

\end{abstract}

\section{Introduction}
The use of silicon detectors in high radiation environments, as to be expected in future 
high energy accelerators or in space missions, poses severe problems due to changes in the 
properties of the material, and consequently influences the performances of detectors.

As a consequence of the degradation to radiation of the semiconductor material, an 
increase of the reverse current due the reduction of the minority carrier lifetime, a 
reduction of the charge collection efficiency and a modification of the effective doping, 
due to the generation of trapping centres, are observed in the detector characteristics.

In this paper, the effects of irradiation conditions and various initial impurities in the 
starting material are discussed in the frame of the phenomenological model able to explain 
quantitatively defect production and evolution toward stable defects during and after 
irradiation in silicon. The model supposes three steps. 

In the first step, the incident particle, having kinetic energies with values in the 
intermediate up to high energy range, interacts with the semiconductor material. The 
peculiarities of the interaction mechanisms are explicitly considered for each kinetic 
energy. 

In the second step, the recoil nuclei resulting from these interactions lose their energy 
in the lattice. Their energy partition between displacements and ionisation is considered 
in accord with the Lindhard theory (\cite{lind} and authors' contributions \cite{anal}).

A point defect in a crystal is an entity that causes an interruption in the lattice 
periodicity. In this paper, the terminology and definitions in agreement with M. Lannoo 
and J. Bourgoin \cite{4} are used in relation to defects. We denote the displacement 
defects, vacancies and interstitials, as primary point defects, prior to any further 
rearrangement. After this step the concentration of primary defects is calculated. 

The mechanisms of interaction of the incident particle with the semiconductor lattice, 
accompanied by displacement defect production have been discussed in some papers, see, 
e.g. references \cite{1, 2, 3}. The incident particle produces, as a consequence of its 
interaction with ions of the semiconductor lattice, cascades of displacements.

In silicon, vacancies and interstitials are essentially unstable and interact via 
migration, recombination, and annihilation or produce other defects.

The concentration of primary defects represents the starting point for the following step 
of the model, the consideration of the annealing processes, treated in the frame of the 
chemical rate theory. A review of previous works about the problem of the annealing of 
radiation induced defects in silicon can be found, e.g. in Reference \cite{5}.

Without free parameters, the model is able to predict the absolute values of the 
concentrations of defects and their time evolution toward stable defects, starting from 
the primary incident particle characterised by type and kinetic energy.

The first two steps have been treated extensively in previous papers \cite{1, 2, 3}, where 
the concentration of primary defects has been calculated. In this paper, the third step is 
discussed extensively and represents a generalisation of the previous results published in 
references \cite{5, 6} including also the carbon contributions to defect kinetics.

The influence and the effects of different initial impurity concentrations of phosphorus, 
oxygen and carbon as well as of the irradiation conditions were systematically studied. 
Vacancy-interstitial annihilation, interstitial migration to sinks, vacancy - impurity 
complexes ($VP$, $V_2$, $VO$, $V_2O$, $C_iO_i$, $C_iC_s$) - only the stable defects 
confirmed experimentally in silicon for high energy physics applications were considered. 
The correlation between the rate of defect productions, the temperature and the time 
evolution of the defect concentrations was also investigated. Some conclusions about the 
possibilities to obtain semiconductor materials harder to radiation are given.

\section{Production of primary defects}
The basic assumption of the present model is that the primary defects, vacancies and 
interstitials, are produced in equal quantities and are uniformly distributed in the 
material bulk. They are produced by the incoming particle, as a consequence of the 
subsequent collisions of the primary recoil in the lattice, or thermally (only Frenkel 
pairs are considered). The generation term $G$ is the sum of two components:
\begin{equation}
G=G_R+G_T
\end{equation}
where $G_R$ accounts for the generation by irradiation, and $G_T$ for thermal generation.
The concentration of the primary radiation induced defects per unit fluence $CPD$ in 
silicon has been calculated as the sum of the concentration of defects resulting from all 
interaction processes, and all characteristic mechanisms corresponding to each interaction 
process, using the explicit formula (see details, e.g. in references \cite{7, 8}):
\begin{equation}
CPD	\left(E\right)= \frac{N_{Si}}{2E_{Si}} \int \sum _{i} \left( \frac{d\sigma}{d\Omega} 
\right)_{i,Si} L(E_{Ri})_{Si} d\Omega=\frac{1}{N_A} \frac{N_{Si}A_{Si}}{2E_{Si}} 
NIEL\left(E\right)
\end{equation}
where $E$ is the kinetic energy of the incident particle, $N_{Si}$ is the atomic density 
in silicon, $A_{Si}$ is the silicon atomic number, $E_{Si}$ - the average threshold energy 
for displacements in the semiconductor, $E_{Ri}$  - the recoil energy of the residual 
nucleus produced in interaction $i$, $L(E_{Ri})$ - the Lindhard factor that describes the 
partition of the recoil energy between ionisation and displacements and 
$(d\sigma/d\Omega)_i$ - the differential cross section of the interaction between the 
incident particle and the nucleus of the lattice for the process or mechanism $i$, 
responsible in defect production. $N_A$ is Avogadro's number. The formula gives also the 
relation with the non ionising energy loss ($NIEL$) the rate of energy loss by 
displacement $(dE/dx)_{ni}$  \cite{4, toti}. 

In Figure 1, the kinetic energy dependence of $CPD$ for different particles is presented: 
for pions the values are calculated in accord with equation (1) and are from reference 
\cite{9}, and for the other particles these are evaluated from the published $NIEL$ 
calculations, as follows: for protons - reference \cite{3, 10}, for neutrons - reference 
\cite{11}, for electrons - reference \cite{3, 10} and for photons from \cite{3}.
The main source of errors in the calculated concentration of primary defects comes from 
the modelling of the particle - nucleus interaction and from the number and quality of the 
experimental data available for these processes. 

\begin{figure}[ht]
\centering
\includegraphics[width=.8\textwidth, clip, angle=90]{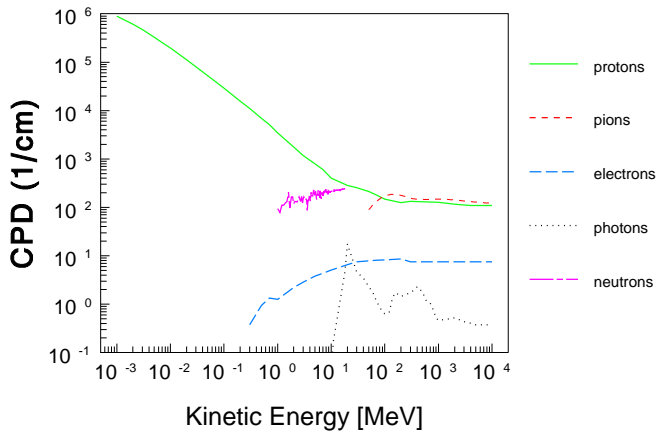}
\caption{\small{Energy dependence of the concentration of primary defects on unit fluence 
induced by protons, pions, electrons, photons and neutrons in silicon - see text for 
details.}}
\label{f1}
\end{figure}

Due to the important weight of annealing processes, as well as to their very short time 
scale, $CPD$ is not a measurable physical quantity. In silicon, vacancies and 
interstitials are essentially unstable and interact via migration, recombination, and 
annihilation or produce other defects.
In the simplifying hypothesis of random distribution of $CPD$ for all particles, used in 
the present paper, the identity of the particle is lost after the primary interaction, and 
two different particles could produce the same generation rate ($G_R$) for 
vacancy-interstitial pairs if the following condition:
\begin{equation}
G_R=[\left(CPD\right)_{part.a}\left(E_1)\right]\cdot\Phi_{part.a}(E_1)=[\left(CPD\right)_{
pa
rt.b}\left(E_2)\right]\cdot\Phi_{part.2(E_2)}
\end{equation}
is fulfilled.
Here, $\Phi$ is the flux of particles ($a$) and ($b$) respectively, and $E_1$ and $E_2$ 
their corresponding kinetic energies.

\section{The kinetics of radiation induced defects}
Silicon used in high energy physics detectors is n-type high resistivity (4 $\div$ 6 
K$\Omega$ cm) phosphorus doped FZ material. 

The effect of oxygen in irradiated silicon has been a subject of intensive studies in 
remote past. In the last decade a lot of studies have been performed to investigate the 
influence of different impurities, especially oxygen and carbon, as possible ways to 
enhance the radiation hardness of silicon for detectors in the future generation of 
experiments in high energy physics - see, e.g. references \cite{12, 13}. Some people 
consider that these impurities added to the silicon bulk modify the formation of 
electrically active defects, thus controlling the macroscopic device parameters. 
Empirically, it is considered that if the silicon is enriched in oxygen, the capture of 
radiation-generated vacancies is favoured by the production of the pseudo-acceptor complex 
vacancy-oxygen. Interstitial oxygen acts as a sink for vacancies, thus reducing the 
probability of formation of the divacancy related complexes, associated with deeper levels 
inside the gap.

The concentrations of interstitial oxygen $O_i$ and substitutional carbon $C_i$ in silicon 
are strongly dependent on the growth technique. In high purity Float Zone Si, oxygen 
interstitial concentrations are around $10^{15}$ cm$^{-3}$, while in Czochralski Si these 
concentrations can reach values as high as $10^{18}$ cm$^{-3}$. Because Czochralski 
silicon is not available in detector grade quality, an oxygenation technique developed at 
BNL produces Diffusion Oxygenated Float Zone in silicon, obtaining a $O_i$  concentration 
of the order 5x10$^{17}$ cm$^{-3}$. These materials can be enriched in substitutional 
carbon up to 1.8x10$^{16}$ cm$^{-3}$.

After the irradiation of silicon, the following stable defects have been identified (see 
References \cite{4, 14}): $Si_i$, $VP$, $VO$, $V_2$, $V_2O$, $C_iO_i$, $C_i$, $C_iC_s$.
The pre-existing thermal defects and those produced by irradiation, as well as the 
impurities, are assumed to be randomly distributed in the solid. An important part of the 
vacancies and interstitials annihilate. The sample contains certain concentrations of 
impurities, which can trap interstitials and vacancies respectively, and form stable 
defects.

Vacancy-interstitial annihilation, interstitial migration to sinks, divacancy, vacancy and 
interstitial impurity complex formation are considered. The role of phosphorus, oxygen and 
carbon is taken into account, and the following stable defects :  $VP$, $VO$, $V_2$, 
$V_2O$, $C_iO_i$, $C_i$, $C_iC_s$ are considered. Other possible defects as $V_3O$, 
$V_2O_2$, $V_3O_3$ \cite{15}, are not included in the present model.

The following picture describes in terms of chemical reactions the mechanisms of 
production and evolution of the defects considered in the present paper:
\begin{equation}                                                        
V+I\ _{\overleftarrow{G}} ^{\underrightarrow{K_1}}\text{annihilation}
\end{equation}
\begin{equation}                                                       
I\stackrel{K_2}{\rightarrow } \text{sinks}
\end{equation}
\begin{equation}							
V+O\ _{\overleftarrow{K_4}} ^{\underrightarrow{K_3}}\ VO
\end{equation}
$VO$ is the $A$ centre.
\begin{equation}							
V+P\ _{\overleftarrow{K_5}} ^{\underrightarrow{K_3}}\ VP
\end{equation}
$VP$ is the $E$ centre.
\begin{equation}						
V+V\ _{\overleftarrow{K_6}} ^{\underrightarrow{K_3}}\ V_2
\end{equation}
\begin{equation}						
V+VO\ _{\overleftarrow{K_{7}}} ^{\underrightarrow{K_{3}}}\ V_2O
\end{equation}
\begin{equation}
I+C_s\stackrel{K_1}{\rightarrow }  C_i
\end{equation}
\begin{equation}
C_i+O_i\stackrel{K_8}{\rightarrow } C_iO_i
\end{equation}
\begin{equation}
A+I\stackrel{K_9}{\rightarrow }  O
\end{equation}
\begin{equation}
C_i+C_s\stackrel{K_8}{\rightarrow }  C_iC_s
\end{equation}
Some considerations about the determination of the reaction constants are given in 
references \cite{5, 6}.

The reaction constants $K_i$ (i = 1, $3 \div 9$) have the general form:
\begin{equation}
K_i=C\nu \exp \left( -E_i/k_BT\right)
\end{equation}
with $\nu$ the vibration frequency of the lattice, $E_i$ the associated activation energy 
and $C$ a numerical constant that accounts for the symmetry of the defect in the lattice.

The reaction constant related to the migration of interstitials to sinks could be 
expressed as:
\begin{equation}
K_2=\alpha \nu \lambda ^2\exp \left( -E_2/k_BT\right)                                                 
\end{equation}
with $\alpha$ the sink concentration and $\lambda$ the jump distance.

The system of coupled differential equations corresponding to the reaction scheme (4) 
$\div$ (13) cannot be solved analytically and a numerical procedure was used.

The following values of the parameters have been used: $E_1$=$E_2$ = 0.4 eV, $E_3$ = 0.8 
eV, $E_4$ = 1.4 eV, $E_5$ = 1.1 eV, $E_6$ = 1.3 eV, $E_7$ = 1.6 eV, $E_8$ = 0.8 eV, $E_9$   
= 1.7 eV, $\nu$ = 10$^{13}$ Hz, $\lambda$ = 10$^{15}$ cm$^2$, $\alpha$ = 10$^{10}$ 
cm$^{-2}$, in accord with standard constants characterising the silicon material, see for 
example the books of Lannoo and Bourgoin \cite{4} or Damask and Dienes \cite{damask}.

\section{Results, discussion, comparison with experimental data and predictions}
The formation and time evolution of stable defects depends on various factors, e.g. the 
concentrations of impurities pre-existent in the sample, the rate of generation, and the 
temperature.

In Figures 2 a $\div$ f, and 3 a $\div$ f the formation and time evolution of the 
vacancy-oxygen, vacancy-phosphorus, divacancy, divacancy-oxygen, carbon interstitial - 
oxygen interstitial and carbon interstitial.- carbon substitutional are modelled in 
silicon containing different initial concentrations of phosphorus, oxygen and carbon, and 
for two very different rates of generation.
\begin{figure}[ht]
\centering
\includegraphics[width=.75\textwidth]{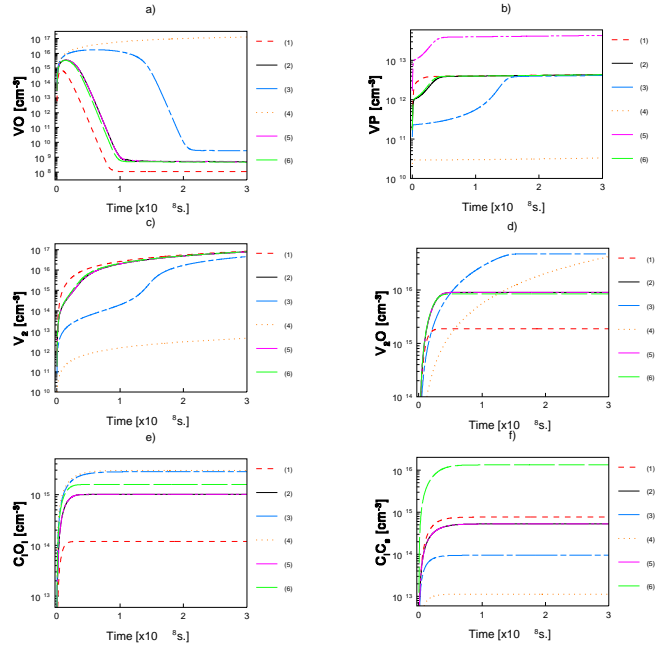}
\caption{\small{Time dependence of the concentrations of a)$VO$, b)$VP$, c)$V_2$, 
d)$V_2O$, e)$C_iO_i$ and f) $C_iC_s$ induced in silicon with the following concentrations 
of impurities: (1): 10$^{14}$ P/cm$^3$, 2x10$^{15}$ O/cm$^3$, and 3x10$^{15}$ C/cm$^3$; 
(2):  10$^{14}$ P/cm$^3$, 10$^{16}$ O/cm$^3$, and 3x10$^{15}$ C/cm$^3$; (3):10$^{14}$ 
P/cm$^3$, 5x10$^{16}$ O/cm$^3$, and 3×10$^{15}$ C/cm$^3$; (4): 10$^{14}$ P/cm$^3$, 
4x10$^{17}$ O/cm$^3$, and 3x10$^{15}$ C/cm$^3$; (5)): 10$^{15}$ P/cm$^3$, 10$^{16}$ 
O/cm$^3$, and 3x10$^{15}$ C/cm$^3$ (6): 10$^{14}$ P/cm$^3$, 10$^{16}$ O/cm$^3$, and 
3x10$^16$ C/cm$^3$by a generation rate $G_R$ = 7x10$^8$ $VI$ pairs/cm$^3$s, at 20$^o$C.}}
\label{f2}
\end{figure}

In Figure 2, the evolution of defect concentrations during high rate irradiation ($G_R$ = 
7x10$^8$ $VI$ pairs/cm$^3$s) is presented. This corresponds, in the model hypothesis, to 
order of magnitude of the radiation levels estimated for the forward tracker region at the 
future LHC accelerator.

The increase of the initial oxygen concentration in silicon, from 2x10$^{15}$ O/cm$^3$ to  
4x10$^{17}$ O/cm$^3$, conduces, after ten years of operation in the field characterised by 
a high and constant generation rate, to the increase of the concentrations of  $VO$ and 
$C_iO_i$ centres, and to the decrease of the concentrations of $V_2$, $VP$ and $C_iC_s$   
ones. With the increase of oxygen concentration, a variation of the $V_2O$ generation rate 
is observed, so that, from the studied cases, the maximum concentration for this defect is 
obtained for 5x10$^{16}$ O/cm$^3$ initial oxygen. The increase of initial phosphorus is 
seen in the increase of concentration of $VP$  centres, while the increase of initial 
carbon concentration has important consequences on the concentrations of $C_iC_s$  
centres. It is interesting to observe that in almost all cases, an equilibrium in reached 
between generation and annealing, and a plateau is obtained in the time dependence of the 
concentrations. The slowest is, in this respect, $V_2O$, that has the highest binding 
energy.

As underlined above, vacancy-oxygen formation in oxygen enriched silicon is favoured in 
respect to the generation of $V_2$, $V_2O$ and $VP$. This is an important feature that 
could be used for detector applications, determining the decrease of the leakage current 
\cite{6}. At high oxygen concentrations, the concentration of $VO$ centres attains a 
plateau during the 10 years period considered. 
\begin{figure}[ht]
\centering
\includegraphics[width=.8\textwidth]{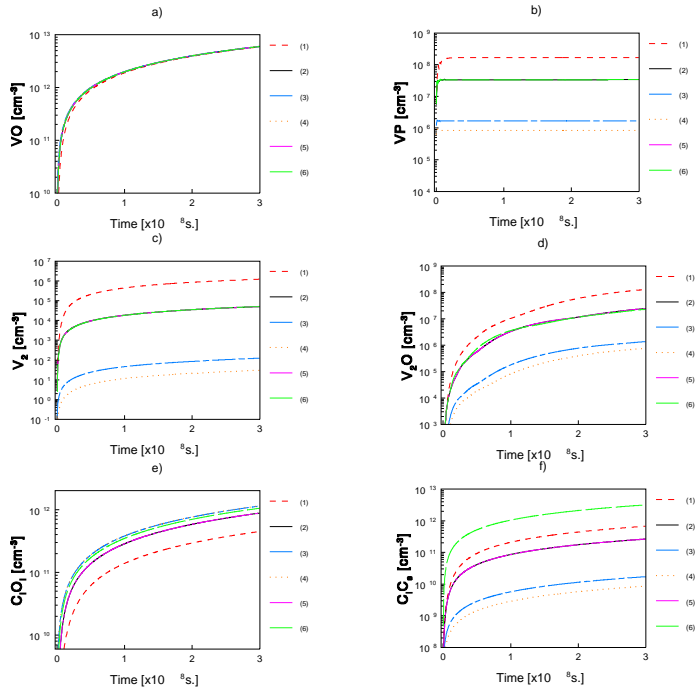}
\caption{\small{Same as Figure 2, $G_R$ = 200 $VI$ pairs/$cm^3$s.}}
\label{f3}
\end{figure}

The other extreme situation corresponds to a generation rate $G_R$ = 200 $VI$ 
pairs/cm$^3$s, equivalent with a rate of defect production by protons from the cosmic ray 
spectra in the orbit near the Earth, at about 400 Km, as will be the position of the 
International Space Station. The same concentrations of $P$, $O$ and $C$ have been 
considered as pre-existent in silicon as in Figure 2. For this generation rate, the 
increase of the oxygen concentration produces the decrease of the concentration of all 
centres, with the exception of the $VO$ concentration, that, at these rates, it is not 
influenced by the oxygen content, and of the $C_iO_i$ concentration, where an increase is 
observed. As could be seen from Figure 3, as a consequence of the small rate of generation 
rate of vacancy - interstitial pairs, after ten years of operation the equilibrium between 
generation and annealing is not reached, the concentrations of defects being, with the 
exception of $VP$ (that has a relatively low binding energy), slightly increasing 
functions of time.

All curves, both from Fig. 2 and Fig. 3, have been calculated for 20$^o$C temperature. 
Thermal generation has been taken into account in both cases, although it is important 
only for the silicon exposed to low rates of defect production.

The influence of the generation rate of primary defects on the concentration of stable 
defects and on their time evolution has also been investigated. In Figure 4 a $\div$ f, 
the time evolution of the $VO$, $VP$, $V_2$, $V_2O$, $C_iO_i$ and $C_iC_s$ concentrations 
is presented for six decades of generation rates of defects ($G_R$=7x10$^5 \div $ 
7x10$^{10}$ $VI$ pairs/cm$^3s$), for silicon containing the following concentrations of 
impurities: 10$^{14}$ P/cm$^3$, 10$^{16}$ O/cm$^3$, and 3×10$^{15}$ C/cm$^3$ at  20$^o$C 
temperature. At small times, the curves corresponding to different generation rates are 
all parallel and equidistant in a log-log representation. Starting from the highest 
generation rates, they start to increase slower ($V_2$, $VP$), attain a plateau $V_2O$, or 
event start to decrease  ($VO$, $C_iO_i$, $C_iC_s$. The maximum attained can be the same, 
independent on the generation rate as is the case of $VO$ concentration, or could be 
generation dependent $C_iO_i$, $C_iC_s$.
\begin{figure}[ht]
\centering
\includegraphics[width=.8\textwidth]{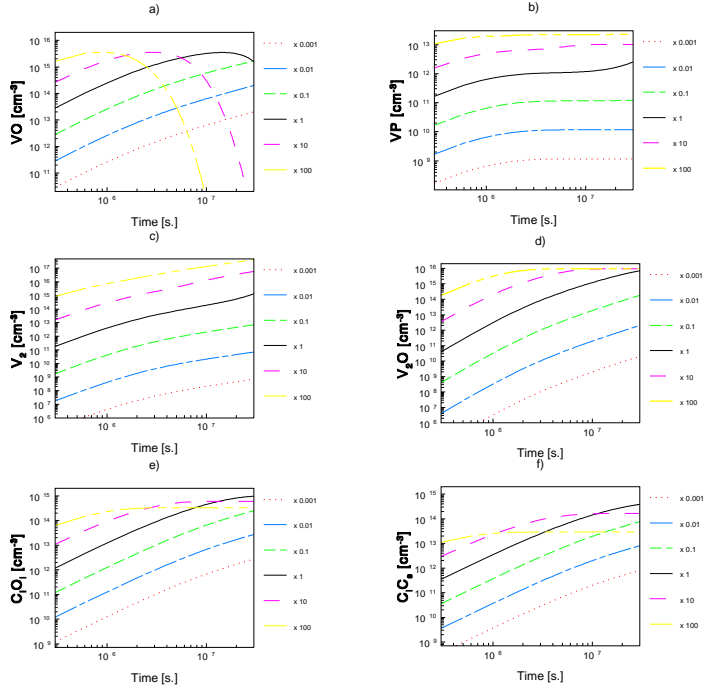}
\caption{\small{Time dependence of the concentrations of a)$VO$, b)$VP$, c)$V_2$, 
d)$V_2O$, e)$C_iO_i$ and f) $C_iC_s$ induced in silicon with:  10$^{14}$ P/cm$^3$, 
10$^{16}$ O/cm$^3$, and 3x10$^{15}$ C/cm$^3$ by continuous irradiation.}}
\label{f4}
\end{figure}

The temperature is another important factor determining the time evolution of defects. In 
Figure 5 a $\div$ f and 6 a $\div$ f, the effect of the temperature is studied for the 
same generation rates of primary defects as in Figures 2 and 3 respectively, for five 
temperatures: 20$^o$C, 10$^o$C, 0$^o$C, -10$^o$C, and -20$^o$C. The concentrations of 
pre-existing impurities in silicon are as follows: 10$^{14}$ P/cm$^3$, 10$^{16}$ O/cm$^3$, 
and 3x10$^{15}$ C/cm$^3$. The decrease of the temperature decreases the generation rate of 
stable defects, with the exception of $VP$, where the highest values correspond to the 
lowest temperature. The maximum values of the $VO$  concentration are temperature 
independent; the values of the concentration on the plateau decrease with the decrease of 
temperature for $C_iO_i$ and $C_iC_s$ and increase for $V_2O$.
\begin{figure}[ht]
\centering
\includegraphics[width=.8\textwidth]{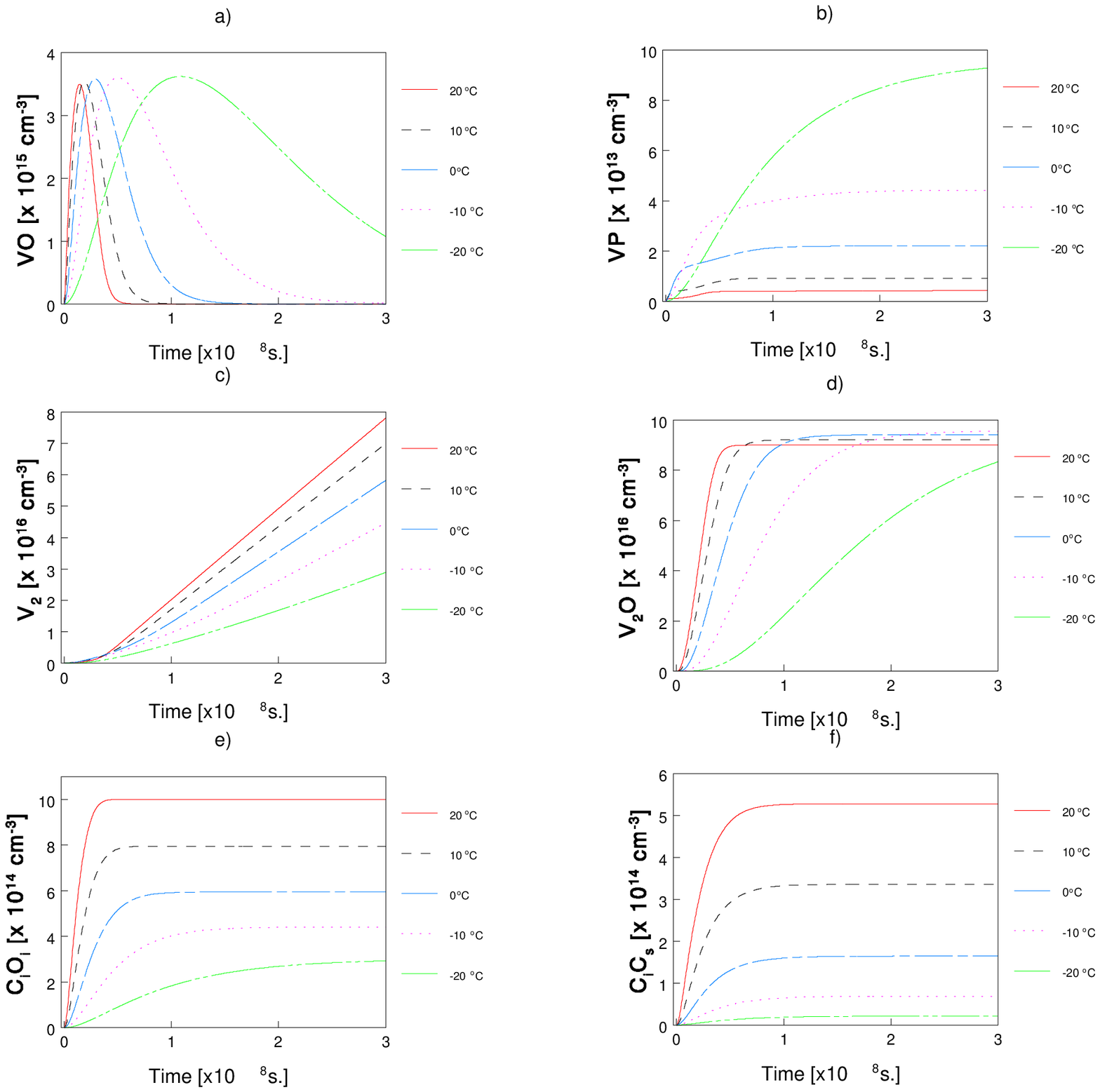}
\caption{\small{Time dependence of the concentrations of a)$VO$, b)$VP$, c)$V_2$, 
d)$V_2O$, e)$C_iO_i$ and f) $C_iC_s$ induced in silicon with:  10$^{14}$ P/cm$^3$, 
10$^{16}$ O/cm$^3$, and 3x10$^{15}$ C/cm$^3$ by continuous irradiation, in the same 
conditions as in Figure 2, at 20$^o$C, 10$^o$C, 0$^o$C, -10$^o$C, and -20$^o$C.}}
\label{f5}
\end{figure}

For silicon exposed to low rates of defect production the same amount of time (Figure 6), 
the process of defect production in slowed down. The most unexpected time dependence is 
for $V_2$, that has the highest values at the lowest temperature.
\begin{figure}[ht]
\centering
\includegraphics[width=.8\textwidth]{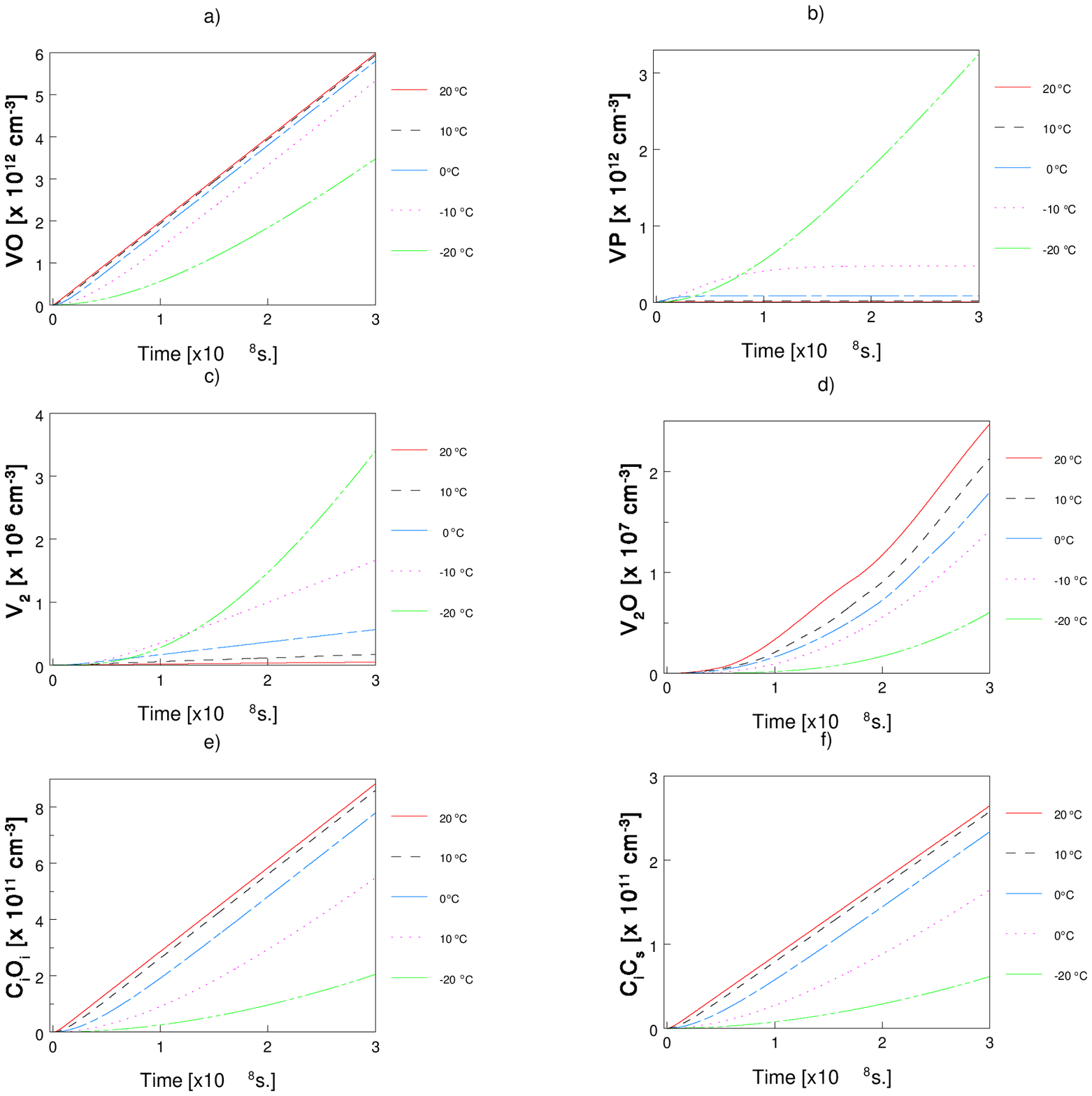}
\caption{\small{Time dependence of the concentrations of a)$VO$, b)$VP$, c)$V_2$, 
d)$V_2O$, e)$C_iO_i$ and f) $C_iC_s$ induced in silicon with:  10$^{14}$ P/cm$^3$, 
10$^{16}$ O/cm$^3$, and 3x10$^{15}$ C/cm$^3$ by continuous irradiation, in the same 
conditions as in Figure 3, at 20$^o$C, 10$^o$C, 0$^o$C, -10$^o$C, and -20$^o$C.}}
\label{f6}
\end{figure}

In a previous paper \cite{6}, we demonstrated in concrete cases the importance of the 
sequence of irradiation process, considering that the same total fluence can be attained 
in different situations: the ideal case of instantaneous irradiation, irradiation in a 
single pulse followed by relaxation, and respectively continuos irradiation process. As 
expected, after instantaneous irradiation the concentrations of defects are higher in 
respect with "gradual" irradiation.

After this analysis, the specific importance of the irradiation and annealing history 
(initial material parameters, type of irradiation particles, energetic source spectra, 
flux, irradiation temperature, measurement temperature, temperature and time between 
irradiation and measurement) on defect evolution must be to underline.

The model predictions have been compared with experimental measurements. A difficulty in 
this comparison is the insufficient information in published papers regarding the 
characterisation of silicon, and on the irradiation parameters and conditions for most of 
the data.

It was underlined in the literature \cite{16} that the ratio of $VO$ to $VP$  centres in 
electron irradiated silicon is proportional to the ratio between the concentrations of 
oxygen and phosphorus in the sample. For electron irradiation, in Ref. \cite{17} a linear 
dependence of the $V_2$  versus $VO$  centre concentration has been found experimentally. 
In the present paper, the ratio of concentrations of  $V_2$ to $VO$  centres and $VO$  to   
$VP$ ones has been calculated in the frame of the model, for the material with the 
characteristics specified in Ref. \cite{17}, and irradiated with 12 MeV  electrons, in the 
conditions of the above mentioned article. The time dependence of these two ratios is 
represented in Figure 7. Annealing is considered both during and after irradiation. It 
could be seen that for the ratio of $V_2$  and $VO$  concentrations the curves 
corresponding to different irradiation fluences are parallel, while the ratio of $VO$  to   
$VP$ concentrations is fluence independent, in the interval 2x10$^{13} \div$ 5.5x10$^{14}$ 
cm$^{-2}$, in good agreement with the experimental evidence. The ratio between $V_2$  and 
$VO$  concentrations is determined by the generation of primary defects by irradiation, 
while the ratio between $VO$  and  $VP$ concentrations is determined by the concentrations 
of oxygen and phosphorus in silicon. 
\begin{figure}[ht]
\centering
\includegraphics[width=.6\textwidth]{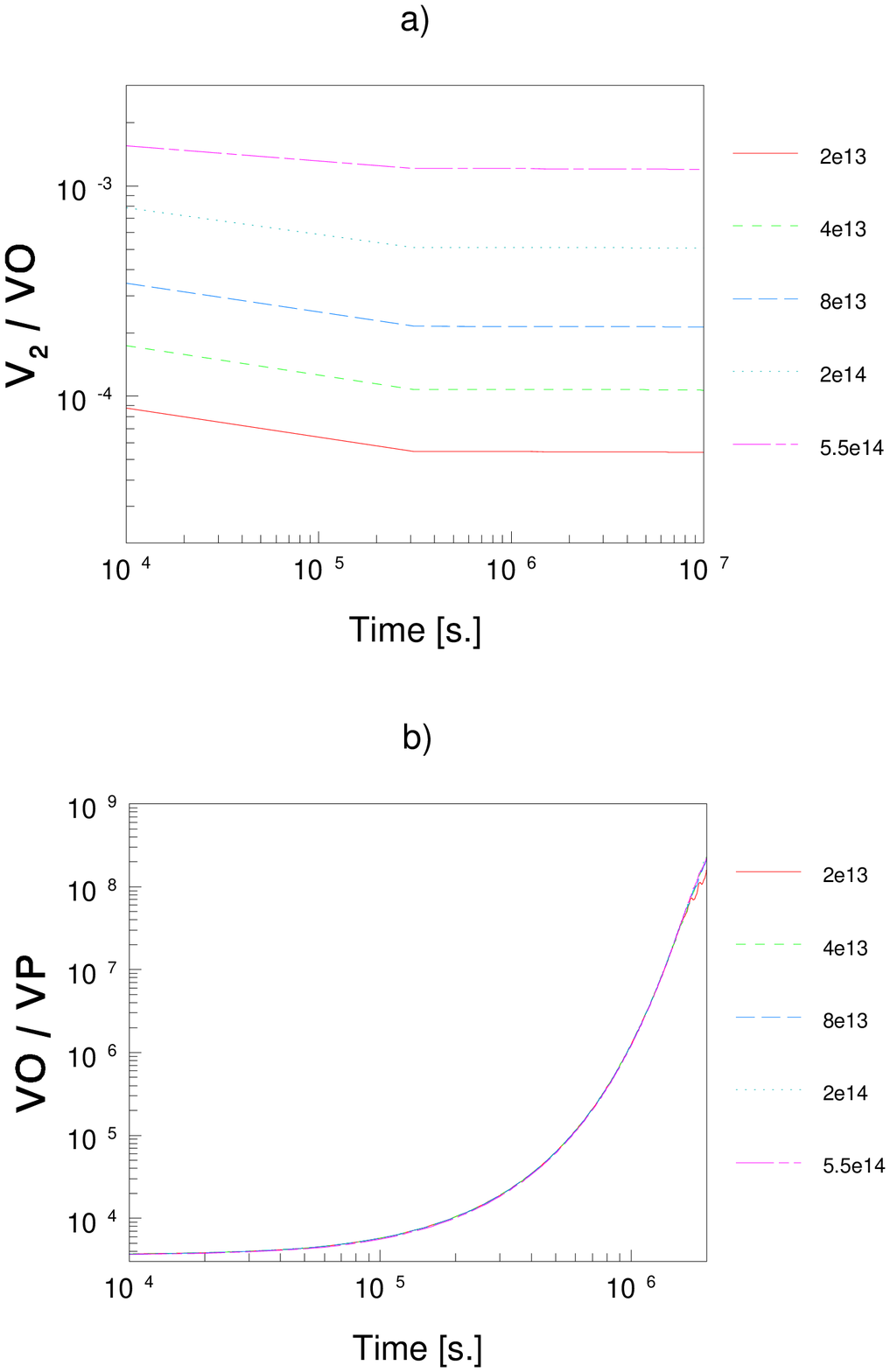}
\caption{\small{Time dependence for a) $V_2/VO$  and b) $VO/VP$  concentrations calculated 
for silicon with 1.4x10$^{14}$ P/cm$^3$, 5x10$^{17}$ O/cm$^3$, and 3x10$^{15}$ C/cm$^3$  
by 12 MeV electron irradiation, with the flux 5.8x10$^{10}$ e/cm$^2$s, up to the fluences: 
2x10$^{13}$ e/cm$^2$, 4x10$^{13}$ e/cm$^2$, 8x10$^{13}$ e/cm$^2$, 2x10$^{14}$ e/cm$^2$  
and 5.5x10$^{14}$ e/cm$^2$, followed by relaxation (see reference \cite{17}.}}
\label{f7}
\end{figure}

Our estimations are also in agreement with the measurements presented in reference 
\cite{18}, after electron irradiation, where defect concentrations are presented as a 
function of the time after irradiation. In Figure 8, both measured and calculated 
dependencies of the $VO$  and $VP$  concentrations are given. The irradiation was 
performed with 2.5 MeV  electrons, up to a fluence of 3x10$^{16}$ e/cm$^2$. A good 
agreement can be observed for these concentrations. The dependencies put in evidence the 
important role played by the carbon-related defects. The relative values are imposed by 
the arbitrary units of experimental data.
\begin{figure}[ht]
\centering
\includegraphics[width=.6\textwidth]{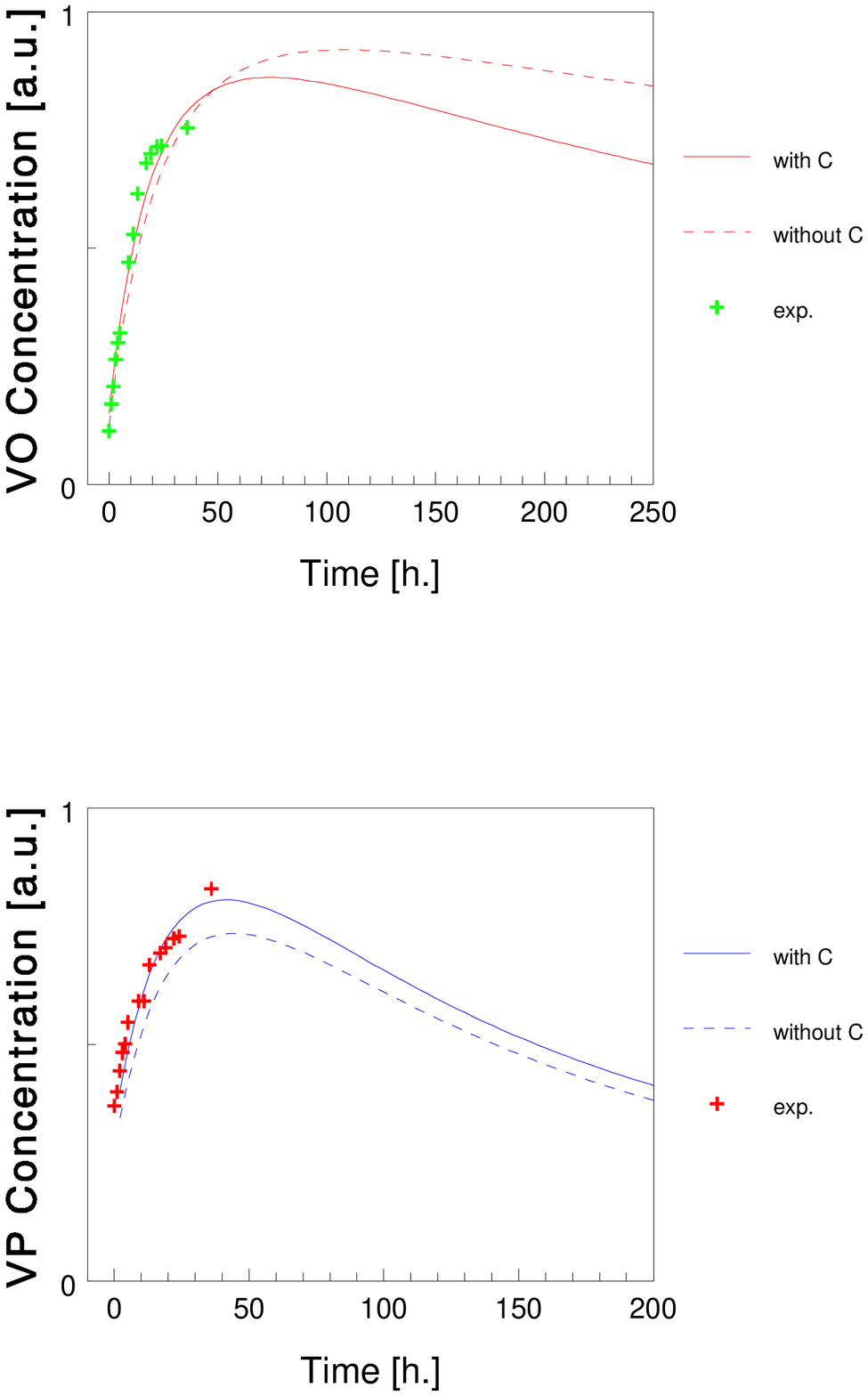}
\caption{\small{Time dependence of $VO$ and $VP$  concentrations after electron 
irradiation: crosses - experimental data from reference \cite{18}]; continuous line - 
present model calculations; dashed line - without the consideration of carbon contribution 
to defect formation.}}
\label{f8}
\end{figure}

Also, a good agreement has been obtained between the absolute values of concentrations of 
$VP+V_2$  and $C_iC_s$ predicted by the model, and the experimental results after   
neutron irradiation at a total fluence of 5.67x10$^{13}$ cm$^{-2}$, reported in reference 
\cite{19}. The calculated 1.5x10$^{13}$ cm$^{-3}$  and 4.1x10$^{12}$ cm$^{-3}$ 
concentrations for $VP+V_2$  and $C_iC_s$ respectively, are in accord with the values of 
1.1x10$^{13}$ cm$^{-3}$ and 3.8x10$^{12}$ cm$^{-3}$, measured experimentally. For the $VO$ 
concentration, a poorer concordance has been obtained.

\section{Summary}
A phenomenological model that describes silicon degradation due to irradiation from the 
point of view of the kinetics of produced defects toward equilibrium was developed. 

The production of primary defects (vacancies and interstitials) in the silicon bulk was 
considered in the frame of the Lindhard theory, and the peculiarities of the particle - 
silicon nuclei interaction were taken into account.

The mechanisms of formation of stable defects and their evolution toward equilibrium was 
modelled, and the concentrations of defects were calculated solving numerically the system 
of coupled differential equations for these processes. Vacancy-interstitial annihilation, 
interstitial migration to sinks, vacancy-impurities complexes ($VP$, $VO$, $V_2O$, 
$C_iO_i$, $C_iC_s$), and divacancy formation were considered in different irradiation 
conditions, for different concentrations of impurities in the initial semiconductor, in 
the temperature range $- 20 \div + 20 ^oC$. 

The calculated results suggest the importance of the conditions of irradiation, 
temperature and annealing history. 

The model supports the experimental studies performed to investigate the influence of 
oxygen in the enhancement of the radiation hardness of silicon for detectors. The  $VO$ 
defects in oxygen enriched silicon are favoured in respect to the other stable defects, 
so, for detector applications it is expected that the leakage current decreases after 
irradiation. At high oxygen concentrations, this defect saturates starting from low 
fluences at high generation rates of defects.

\end{document}